\documentclass[modern]{aastex61}

\submitjournal{the Astrophysical Journal}

\usepackage{graphicx}	
\usepackage{amsmath}	
\usepackage{amssymb}	
\usepackage{bm}		    
\usepackage{mathrsfs}   
\usepackage{natbib}     
\usepackage{bookmark}   
\usepackage{xcolor}
\usepackage{color}
\usepackage[T1]{fontenc}

\usepackage{caption}

\usepackage[final]{microtype}

\usepackage{needspace}

\let\splitbox\relax
\usepackage{adjustbox} 

\newcommand{\du}{\frac{\delta u}{u}}
\newcommand{\dun}{\frac{\delta u'}{u'}}
\newcommand{\dM}{\delta M}
\newcommand{\dR}{\delta R}
\newcommand{\dMM}{\frac{\dM}{M}}
\newcommand{\dRR}{\frac{\dR}{R}}
\newcommand{\dY}{\delta Y}
\newcommand{\KuY}{K_i^{(u',Y)}}
\newcommand{\KYu}{K_i^{(Y,u')}}

\definecolor{echelle-yellow}{HTML}{F29559}
\definecolor{echelle-blue}{HTML}{0571b0}
\definecolor{echelle-red}{HTML}{DB4D48}
\definecolor{echelle-black}{HTML}{323031}

\definecolor{turn-orange}{HTML}{B16D41}

\definecolor{diff-blue}{HTML}{0571b0}
\definecolor{diff-red}{HTML}{ca0020}
\definecolor{diff-orange}{HTML}{f97100}
\definecolor{diff-purple}{HTML}{551a8b}

\makeatletter
\newif\ifhbonecolumn
\@ifclasswith{aastex61}{twocolumn}{\hbonecolumnfalse}{\hbonecolumntrue}
\makeatother

\newcommand{\colwidth}{\ifhbonecolumn 0.5\linewidth\else \linewidth\fi}
\newcommand{\figfactor}{\ifhbonecolumn 0.66 \else 1 \fi}
\newcommand{\captwidth}{\ifhbonecolumn 7in \else\colwidth\fi}
\newcommand{\widefigwidth}{\ifhbonecolumn 1.3\linewidth\else\linewidth\fi}
\newcommand{\thinfigstar}{\ifhbonecolumn 0.66\linewidth\else 0.5\linewidth\fi}

\newcommand{\bottrim}{\ifhbonecolumn 1.4 \else 1.2 \fi}
\newcommand{\righttrim}{\ifhbonecolumn 0.55 \else 0.45 \fi}

\newif\ifref
\reffalse
\newcommand{\mb}[1]{\ifref\boldmath\textbf{#1}\unboldmath\else #1\fi}

\shorttitle{Asteroseismic Inversions for Stellar Structure}
\shortauthors{E.~P.~Bellinger et al.}

\begin{document}

\title{Model-independent measurement of internal~stellar~structure in 16~Cygni~A \& B} 

\correspondingauthor{Earl P.~Bellinger}
\email{bellinger@mps.mpg.de}

\author[0000-0003-4456-4863]{Earl P.~Bellinger,}
\altaffiliation{Fellow of the National Physical Science Consortium}
\affil{Max-Planck-Institut f{\"u}r Sonnensystemforschung, Justus-von-Liebig-Weg 3, 37077 G{\"o}ttingen, Germany}
\affil{Department of Astronomy, Yale University, New Haven, CT 06520, USA}
\affil{Stellar Astrophysics Centre, Department of Physics and Astronomy, Aarhus University, Ny Munkegade 120, DK-8000 Aarhus C, Denmark}
\affil{Institut f{\"u}r Informatik, Georg-August-Universit{\"a}t G{\"o}ttingen, Goldschmidtstrasse 7, 37077 G{\"o}ttingen, Germany}
\collaboration{}

\author{Sarbani Basu}
\affil{Department of Astronomy, Yale University, New Haven, CT 06520, USA}

\author{Saskia Hekker}
\affil{Max-Planck-Institut f{\"u}r Sonnensystemforschung, Justus-von-Liebig-Weg 3, 37077 G{\"o}ttingen, Germany}
\affil{Stellar Astrophysics Centre, Department of Physics and Astronomy, Aarhus University, Ny Munkegade 120, DK-8000 Aarhus C, Denmark}

\author{Warrick Ball}
\affil{School of Physics and Astronomy, University of Birmingham, Edgbaston, Birmingham B15 2TT, United Kingdom}
\affil{Stellar Astrophysics Centre, Department of Physics and Astronomy, Aarhus University, Ny Munkegade 120, DK-8000 Aarhus C, Denmark}

\begin{abstract}
We present a method for measuring internal stellar structure based on asteroseismology that we call ``inversions-for-agreement.'' 
The method accounts for imprecise estimates of stellar mass and radius as well as the relatively limited oscillation mode sets that are available for distant stars. 
\mb{By construction, the results of the method are independent of stellar models.} 
We apply this method to measure the isothermal sound speeds in the cores of the solar-type stars 16~Cyg~A and B using asteroseismic data obtained from \emph{Kepler} observations. 
We compare the asteroseismic structure that we deduce against best-fitting evolutionary models and find that the sound speeds in the cores of these stars exceed those of the models. 
\end{abstract}


\section{Introduction} 
\ifhbonecolumn\doublespace\fi

The detection and study of internal waves in stars---asteroseismology---provides a unique view into stellar interiors. 
As the structure of a star dictates the varieties and frequencies of its normal modes of oscillation, asteroseismic data can be used to set limits on the conditions inside a star. 
This is usually achieved by evolving stellar models, and the structure of the best-fitting model is then assumed to be a proxy for the structure of the star. 
However, theoretical pulsation frequencies of even the best stellar models have significant discrepancies with observations, implying that the structure of the star differs from the structure of the model. 
This is true for the Sun and other stars alike. 
A way to proceed from this point would be to quantify what internal conditions do support the oscillations that have been observed. 
This problem is the inverse of determining the mode frequencies of a known stellar structure, and is thus known as a \emph{structure inversion}. 
Structure inversions are of value because their results are independent of models. 
However, the structure inversion problem is ill-posed in the sense described by \citet{hadamard} and therefore difficult to solve, especially given the relatively limited data that are available for other stars. 
Consequently, structure inversions for internal properties such as the sound-speed profile have thus far been restricted to the Sun and other bodies within the solar system. 
In this paper, we present results of structure inversions performed to probe core structure in other stars. 
More specifically, we invert measured p-mode frequencies to deduce the squared isothermal sound speed ($u\equiv P/\rho$, where $P$ is pressure and $\rho$ is density), in the cores of the two solar analogs 16~Cyg~A and 16~Cyg~B. 
We achieve this by introducing an algorithm that we call ``inversions-for-agreement'' that works with the available data.

Helioseismic inversions, i.e.~inversions for the Sun, have revealed that sound-speed profiles of solar-calibrated evolutionary models differ by only fractions of a percent from the actual structure of the Sun---a rare triumph of accuracy by astrophysical standards. 
Furthermore, even before all flavors of solar neutrinos could be detected, helioseismic inversions were instrumental in showing that the solar neutrino problem was external to solar modelling \citep[e.g.][]{1997MNRAS.289L...1A,1998PhLB..433....1B}.
Additionally, the importance of some physical processes in stellar physics have been revealed by helioseismic inversions as well. 
For example, by comparing solar models with and without diffusion and gravitational settling of helium and heavy elements, \citet{1993ApJ...403L..75C} showed that it is important to take these effects into account (see also Figure~20 of \citealt{2016LRSP...13....2B}), and it has now become common practice to include these processes when modelling other solar-like stars. 
Hence, structure inversions are useful for verifying and improving models both within stellar physics and beyond.

\begin{figure*}
    \centering
    \captionsetup{width=\captwidth,font=small}
    \makebox[\linewidth][c]{
        \includegraphics[width=\widefigwidth]{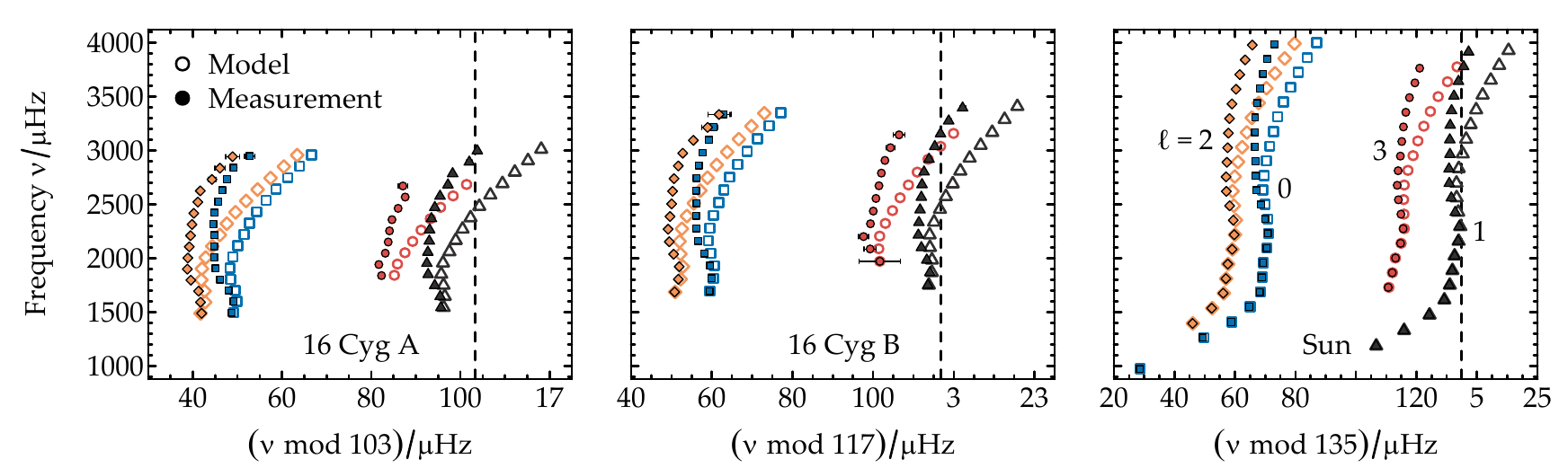}
    }
    \caption{\'Echelle diagrams comparing \emph{GOE} evolutionary models of 16~Cyg~A (left) and B (center) to frequencies extracted from \emph{Kepler} data. 
    For reference, the right panel shows the solar model Model S \citep{1996Sci...272.1286C} in comparison with low-degree frequencies of the quiet Sun from BiSON data \citep[][]{2014MNRAS.439.2025D}. 
    The dashed line indicates the large frequency separation ($\Delta\nu$). 
    Open symbols are model frequencies and filled symbols are observed frequencies. 
    Spherical degrees $\ell$ are indicated with color and shape: 
    \textcolor{echelle-blue}{0} (blue squares), 
    \textcolor{echelle-black}{1} (black triangles), 
    \textcolor{echelle-yellow}{2} (yellow diamonds), and
    \textcolor{echelle-red}{3} (red circles). 
    Error bars show $1\sigma$ uncertainties, which in most cases are not visible. Model frequencies significantly differ from observed frequencies in nearly all cases. } 
    \label{fig:echelle} 
\end{figure*}

The stars we wish to study with structure inversions are pulsating solar-type stars observed by \emph{Kepler}. 
They are cool dwarf stars on the main sequence that pulsate in pure p-modes and show no signs of mode mixing \citep[for a review of solar-like oscillations, see e.g.][]{2013ARA&A..51..353C}. 
The precise measurement of pulsation frequencies in these and other similar stars has enabled estimates of their ages, masses, and radii to better than 15\%, 4\%, and 2\%, respectively \citep{2015MNRAS.452.2127S, 2017ApJ...835..173S, 2016ApJ...830...31B, 2017arXiv170506759B, 2017ApJ...839..116A}. 
The solar-type stars belonging to the triple system of 16~Cygni are two of the most well-studied stars in this field. 
Though stellar models of these stars match the overall characteristics of the stars, such as their radii, luminosities, temperatures, and metallicities; an inspection of their mode frequencies reveals significant disagreements. 
Figure~\ref{fig:echelle} shows a comparison of mode frequencies between models \citep[][models \emph{GOE}]{2017ApJ...835..173S} and observations \citep{2015MNRAS.446.2959D} of 16~Cyg~A and B, with Sun-as-a-star data shown for reference. 
Clear differences can be seen between the mode frequencies of the evolutionary models and the measured mode frequencies of the stars.

\begin{figure*}
    \centering
    \captionsetup{width=\captwidth,font=small}
    \makebox[\linewidth][c]{%
        \includegraphics[width=\widefigwidth]{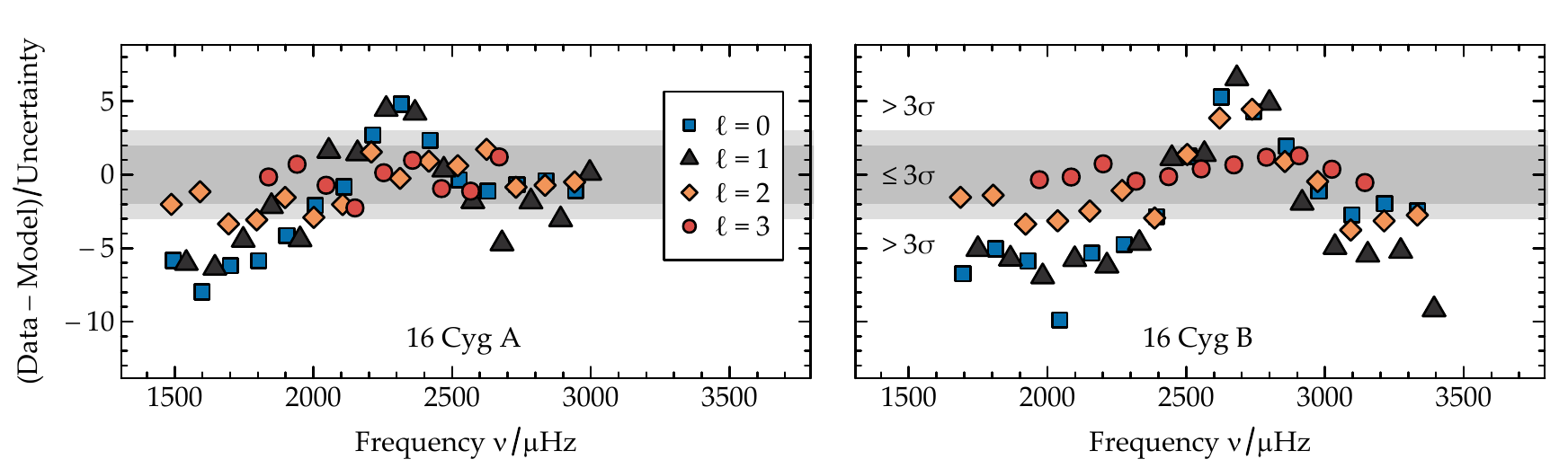}
    }
    \caption{Differences in oscillation mode frequencies between models and observations after correcting for surface effects. 
    Mode frequencies that lie outside of the shaded regions, demarcating the $2\sigma$ and $3\sigma$ boundaries, have significant differences that are caused by differences in internal structure. 
    \label{fig:bg-corr}} 
\end{figure*}

The most conspicuous difference between the oscillations of stars and stellar models is an offset that increases with frequency. This offset arises due to inadequacies in modelling the effects of convection in the near-surface layers \citep[see e.g.,][]{1984srps.conf...11C} as well as neglected treatment of pulsation-convection interaction \citep{2017MNRAS.464L.124H}. 
These are collectively known as ``surface effects,'' and the offset they produce is usually called the ``surface term.'' 
For modes of low spherical degree $\ell$, the surface term is a function of frequency alone. 
There are a number of methods for correcting the disparities imposed by surface effects, such as those given by \citet{2008ApJ...683L.175K}, \citet[][hereinafter BG14]{2014A&A...568A.123B}, and \citet{2015A&A...583A.112S}. 
Each of these methods work by assuming that the frequency offset due to the surface term has a particular form that can be fitted to the frequency differences and subtracted off. 
Even after correction for the surface term, however, differences remain. 
Figure~\ref{fig:bg-corr} shows the remaining discrepancies between mode frequencies of models and observations of 16~Cygni after subtracting off the two-term ``BG14-2'' surface effect. 
More than half of the surface-term corrected mode frequencies still have significant differences with the observed values. 
Moreover, the disparities are most significant in the radial and dipole modes, which probe the deep interior of the star. 

Since mode frequencies of models produced by stellar evolution codes have significant differences with respect to observations even after correction for the surface term, we pursue the use of inversion techniques to make more direct determinations of stellar structure.

\Needspace{3\baselineskip}
\subsection{The inversion problem}
Structure inversions can be posed as the problem of deducing small differences in structure between a star and a sufficiently close reference model by comparison of their mode frequencies. 
The basic problem is the same as the structure inversion problem for the Sun \citep[for reviews of solar structure inversions, see for example][]{Kosovichev1999, 2016LRSP...13....2B}. 
The dependence of mode frequencies on the radial structure of a star is non-linear and involves unobservable displacement eigenfunctions. 
However, the oscillation equations are, to first order, a set of Hermitian eigenvalue equations \citep{1964ApJ...139..664C}, and hence they can be linearized around a known model using the variational principle. 
The linearization links the differences in frequencies between the reference model and the star to the differences in their internal structure. 
A byproduct of the linearization is the fact that the differences must be considered with respect to at least two stellar structure functions simultaneously, as variables such as the sound~speed~$c$ and density~$\rho$ are not independent but rather related through the equations of stellar structure. 
The equations resulting from the linearization can be written as: 
\begin{equation} \label{eq:inversion}
    \mathscr{P} [\nu_i] = \int \mathbf{K}_i(r) \cdot \mathscr{P} [\boldsymbol{f}(r)] \; \text{d}r + \epsilon_i, \quad i \in \mathscr{M}
\end{equation}
where $\mathscr{M}$~is the set of observed modes, 
$\boldsymbol{\nu}$~are the oscillation frequencies of those modes, 
$\boldsymbol{f}$~contains two stellar structure functions (i.e., $f_1(r)$ and $f_2(r)$; e.g.~$c(r)$ and $\rho(r)$), 
$r$~is the fractional stellar radius, and 
$\mathscr{P}$~is a perturbation operator (in this case, the relative difference operator). 
Since measurements are uncertain, we include a term $\boldsymbol \epsilon$ for the differences between the true and the measured values. 
Each mode of oscillation $i$ has its own pair of kernels $\mathbf {K}_i$ that relate changes in $\mathbf f$ to changes in $\nu_i$. 
The kernels are derived from the perturbation analysis (see e.g.~\citealt{GoughThompson1991} or Sec.~6.2.~of \citealt{2016LRSP...13....2B} for details) and can be computed for a given reference model. 
Since the eigenproblem is Hermitian, perturbations to the oscillation mode eigenfrequencies do not depend to the first order on perturbations to the mode eigenfunctions. 
The inverse problem is thus to deduce $\mathbf f$ from the data $\boldsymbol \nu$, given that the kernels are known. 
There is no analytic solution to this problem and numerical methods must be employed. 
In practice, another term must also be added in order to account for the aforementioned surface effects. 
Although the technique makes use of a reference model, the results are independent; all stellar models within the linear regime produce essentially the same inference about the star \citep{2000ApJ...529.1084B}. 
We expand Equation (\ref{eq:inversion}) explicitly in the next section.

Like many inverse problems, the structure inversion problem is ill-posed: the solutions are not unique, and they are also unstable with respect to small fluctuations in the oscillation data (see \citealt{GoughThompson1991} for a discussion). 
Solutions must therefore be regularized \citep[for a review of statistical regularization, see e.g.][]{10.2307/3649759}. 
There are two popular ways of inverting Equation (\ref{eq:inversion}): the Regularized Least Squares \citep[RLS,][]{tikhonov1977solutions} fitting method, which attempts to determine the stellar structure functions $\mathbf f$ that best fit to the observed data; and (2) the method of Optimally Localized Averages \citep[OLA,][]{1968GeoJ...16..169B}, which attempts to make linear combinations of the data that correspond to localized averages of one of the two components of $\mathbf f$. 
Both methods have been used extensively in the case of the Sun.
Details of how the inversions are implemented can be found in \citealt{2016LRSP...13....2B} and references therein.

\begin{figure*}
    \centering
    \captionsetup{width=\captwidth,font=small}
    \makebox[\colwidth][c]{%
        \adjustbox{trim={0.02\width} 0.005cm {0.04\width} 0.01cm, clip}{
            \includegraphics[width=\thinfigstar]{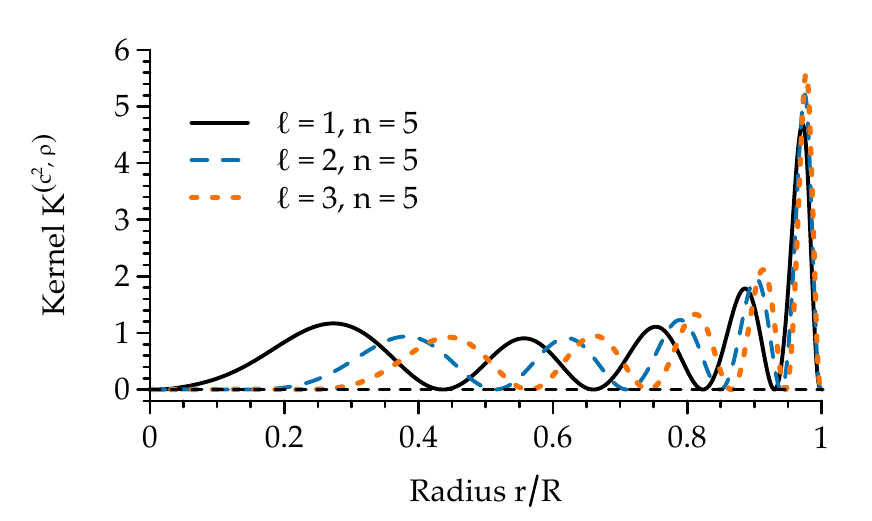}
        }
        \adjustbox{trim={0.04\width} 0.005cm {0.02\width} 0.01cm, clip}{
            \includegraphics[width=\thinfigstar]{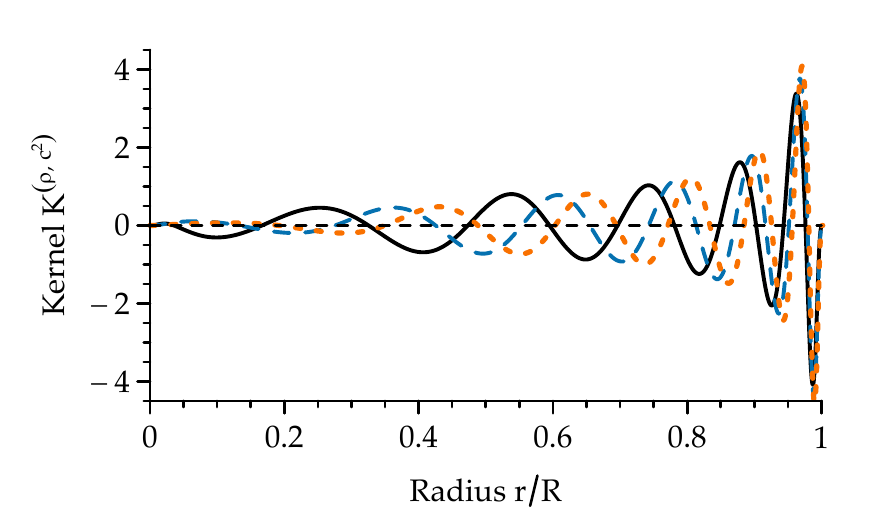}}
        }
    \caption{Kernels for the squared adiabatic sound speed and density, $K^{(c^2, \rho)}$ (left), and the reverse, $K^{(\rho, c^2)}$ (right), as a function of fractional radius for oscillation modes of model \emph{GOE} of 16~Cyg~A. Kernels are shown for modes with the same radial order $n$ but different spherical degree $\ell$ (see legend). 
    \label{fig:c2-rho} }
\end{figure*}

In helioseismic investigations, the most common choice of $\mathbf f$ is the combination of squared adiabatic sound speed $c^2$ and density $\rho$.
The kernels for this pair are shown in Figure~\ref{fig:c2-rho}. 
The basic ingredients of helioseismic inversion are the thousands of precisely measured solar mode frequencies whose spherical degrees range up to $\ell \simeq 200$ or higher. 
Reference models have the same mass, radius, and age as the Sun. 
Inversion of helioseismic data yields inferences of solar structure throughout most of the solar interior \citep[see e.g.][]{2009ApJ...699.1403B}.

There are two major difficulties in trying to invert for the structure of other stars.
The first difficulty is the lack of data. 
Even for the best solar-type targets, only about 55 mode frequencies have been able to be measured. 
Furthermore, due to cancellation effects, we only get data for low-degree modes, usually of degree $\ell=0$, 1, 2, and sometimes 3. 
This limits the regions in the star that we are able to probe, the inversion techniques that we are able to employ, and the pair of stellar structure functions that we are able to use. 
Second, when compared with the Sun, masses and radii of stars are not known with the same precision. 
This is problematic because differences in mass and radius between the reference model and the proxy star cause systematic errors in the inversion results \citep[see][]{2003Ap&SS.284..153B}. 
Most of the time, these quantities are not known independently and need to be determined from the same set of data. 
Even where independent estimates are available, such as radii from interferometric measurements, the uncertainties are non-negligible. 
Both the amount of data and the precision to which the stellar mass and radius are known cause difficulties in inversion of asteroseismic data, and therefore the inversion methods need to be modified. 

\Needspace{3\baselineskip}
\subsection{Asteroseismic inversions} 
\label{sec:asteroseismic-inversions}

Even before CoRoT and {\it Kepler} detected oscillations in a large number of stars, there were a number of studies that investigated the possibility of inverting asteroseismic p-mode oscillations to determine the core structures of solar like stars \citep{1993ASPC...40..541G,
1998mons.proc...33G, 2001ESASP.464..411B, 2001ESASP.464..407B,
2002ESASP.485..249B, 2003Ap&SS.284..153B}. 
Additionally, there was at least one inconclusive study that tried to perform an inversion of seismic data from Procyon~A \citep{2004ESASP.559..186D}. 
The theoretical investigations of structure inversions all used mode sets and data uncertainties that were expected to be available from future missions to determine how well the structure differences between the cores of pairs of models could be determined. 
Unfortunately, the assumptions about the available mode sets and uncertainties were rather optimistic when compared with data available today.

\Needspace{3\baselineskip}
\subsubsection{Mode Set}
\begin{figure}
    \centering
    \captionsetup{width=\captwidth,font=small}
    \makebox[\linewidth][c]{%
    \includegraphics[width=\figfactor\linewidth]{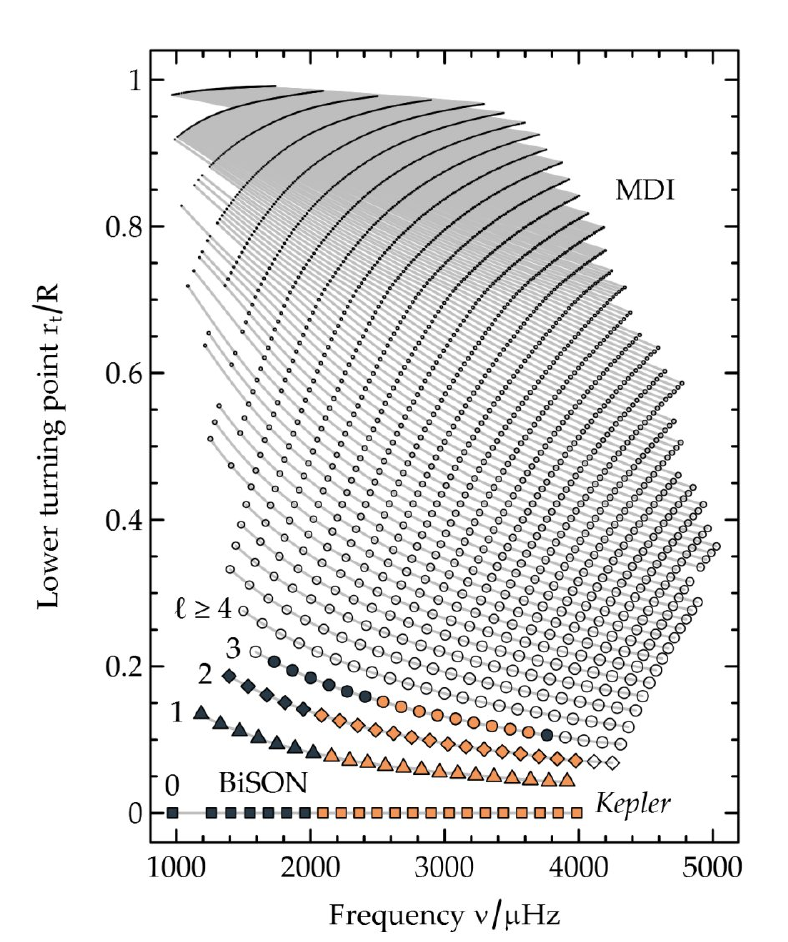}
    }
    \caption{Lower turning points as a function of frequency for oscillation modes of a solar model with the MDI mode set (all points), BiSON mode set (all filled points) and the 16~Cyg~A mode set \mb{from \emph{Kepler}} (\textcolor{turn-orange}{orange} filled points). 
    Modes of the same spherical degree are connected by lines, with modes of spherical degree $\ell = 0$, 1, 2, and 3 shown with squares, triangles, diamonds, and circles, respectively. Compared to the Sun, asteroseismology of solar-like oscillators is restricted to low-degree, high-frequency modes. 
    \label{fig:turning-points}} 
\end{figure}

The limited mode set available for stars other than the Sun makes the inversion problem more difficult. 
The fact that we cannot make resolved-disk observations of other stars generally restricts the detection of modes to $\ell \le 3$. 
The lower turning points of these modes are within the stellar core; consequently, lacking more shallowly-trapped modes, we will be unable to resolve the details of the stellar envelope. 
Figure~\ref{fig:turning-points} illustrates this difficulty by comparing the propagation cavities of oscillation modes with different degrees from a solar model. 
The figure shows lower turning points for low-degree Sun-as-a-star modes obtained by the Birmingham Solar Oscillation Network \citep[BiSON,][]{2014MNRAS.439.2025D} and the $\ell > 3$ modes obtained by the Michaelson Doppler Imager (MDI) mission on board the Solar and Heliospheric Observatory \citep[SOHO,][]{1997SoPh..175..287R}. 
The figure further shows the mode set that would be available if the Sun were a star in the {\it Kepler} field. 
Such a restricted mode set eliminates the possibility of using an inversion technique such as RLS that requires simultaneous determination of $f_1$ and $f_2$ over as large a part of the star as possible. 
Instead, we are confined to investigations of the stellar core.

\begin{figure*}
    \centering
    \captionsetup{width=\captwidth,font=small}
    \makebox[\colwidth][c]{%
        \adjustbox{trim={0.02\width} 0.005cm {0.04\width} 0.01cm, clip}{
            \includegraphics[width=\thinfigstar]{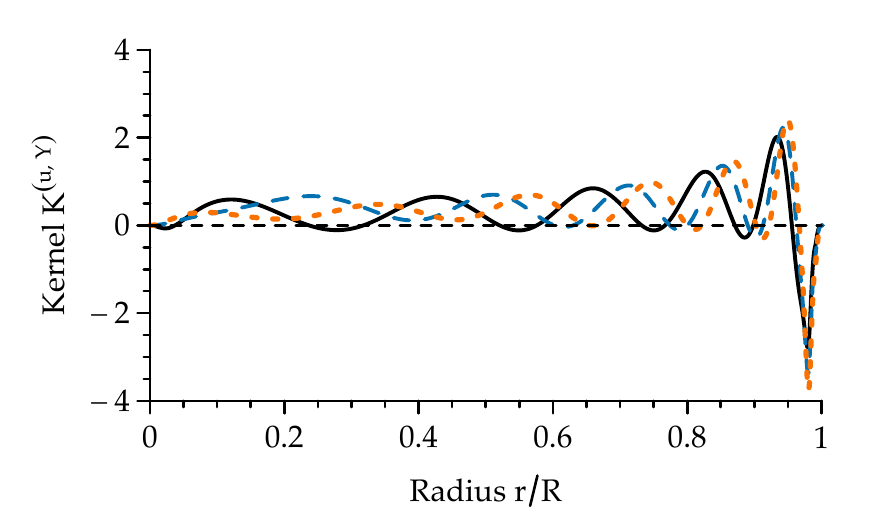}
        }
        \adjustbox{trim={0.04\width} 0.005cm {0.02\width} 0.01cm, clip}{
            \includegraphics[width=\thinfigstar]{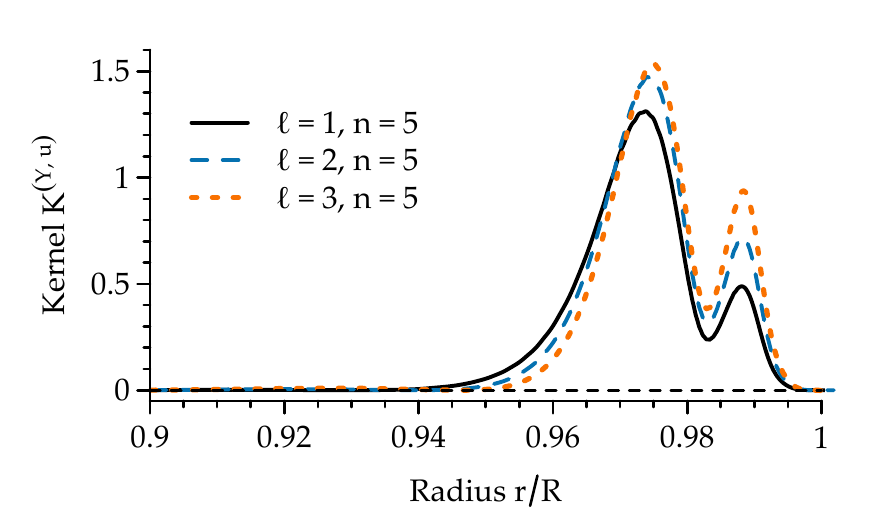}}
        }
    \caption{Kernels for the squared isothermal sound speed and helium abundance, $K^{(u, Y)}$ (left), and the reverse, $K^{(Y, u)}$ (right), as a function of fractional radius for oscillation modes of model \emph{GOE} of 16~Cyg~A. Notice that in contrast to the $K^{(\rho, c^2)}$ kernels shown in Figure~\ref{fig:c2-rho}, the $K^{(Y, u)}$ kernels have very small values ($0 < K(r) < 0.01$) in the interior $r < 0.9\; R$. \label{fig:same-n-uY} }
\end{figure*}

Inversions using the OLA method or its variants are most suited for asteroseismic inversions, since OLA allows inversions over a small part of the star. 
\citet{2003Ap&SS.284..153B} showed that instead of the $(c^2,\rho)$ pair of variables used in solar inversions, the $(u, Y)$ pair is better suited for asteroseismic structure inversions\mb{, where $Y$ is the fractional helium abundance}. 
This is because the kernels for $Y$ are non-zero only in the helium ionization zone, as shown in Figure~\ref{fig:same-n-uY}. 
Thus from the point of view of Equation (\ref{eq:inversion}) the data, i.e., the frequency differences, are almost completely determined by differences in $u$, thereby making $u$ easier to determine. 
However, in order to derive the kernels for the $(u, Y)$ pair, we have to assume that the EOS of the star is the same as that of the reference model \citep{1990MNRAS.244..542D, Kosovichev1999, ThompsonJCD2002}. 
In other words, we are artificially adding information to the system. 
\citet{1997A&A...322L...5B} have shown that in the case of the Sun, this results in systematic errors in the inversion result; 
however, for other stars, we expect the errors caused by data uncertainties to be much larger than the systematic errors caused by an incorrect EOS. 
Thus, we proceed with this pair of variables.

\Needspace{3\baselineskip}
\subsubsection{Mass \& Radius}
The reduced precision of mass $M$ and radius $R$ estimates for stars other than the Sun also makes the problem more difficult. 
Frequencies scale as the square root of mean density, i.e., $\nu^2 \propto M/R^3$, so an unaccounted for difference in $M$ and $R$ between the star and the reference model gives rise to additional systematic errors in the inversion result. 
As these errors are proportional to the uncertainties in $M$ and $R$, they are much larger than those expected from an incorrect EOS. 
Solar inversions as well as trial inversions for stellar models have hitherto been performed under the assumption that the mass and radius of the star are known. 
Having imprecise estimates of the stellar mass and radius means that the mass and radius of the reference model are likely to differ from those of the star. 
\citet{2001ESASP.464..411B} accounted for this effect in their tests of asteroseismic inversions with pairs of models by adding terms for $\delta M$ and $\delta R$ to the inversion procedure. 
However, they assumed $\delta M$ and $\delta R$ to be known exactly, and the impact of uncertainties was not explored in that work.

Another difficulty arises from the fact that the inversion equation and the kernels are usually derived using dimensionless units, with the relative differences in $f_1$ and $f_2$ being calculated at constant fractional radii.
This raises complications alluded to earlier: the $u$ inversion result itself is also systematically offset by the differences in mass and radius \citep{2003Ap&SS.284..153B}.
In short, since kernels are derived using dimensionless variables, instead of a dimensional $u$, we actually have $u' \equiv P'/\rho'$, where $'$ denotes a dimensionless variable. 
It is straightforward to see from the equation governing conservation of mass that $\rho \propto M/R^3$. 
Likewise from the equation of hydrostatic support one finds that $P \propto M^2/R^4$. 
Hence $u'=uR/M$, and so an inversion whose reference model has a different $M$ or $R$ will result in a $u$ profile that differs by
\begin{equation} \label{eq:MR}
    \dun - \du = \dRR - \dMM.
\end{equation}
Thus the inversion procedure must be modified in order to accommodate the reduced precision of mass and radius estimates. 

\vspace{1cm}

These difficulties---limited mode sets and the uncertainties in stellar mass and radius estimates---have so far prevented structure inversions from widespread application in other stars. 
In this paper, we propose a way to circumvent the systematic error that results from the reference model having an incorrect mass and radius by extending the inversion procedure to use multiple reference models spanning the uncertainties in mass and radius. 
Furthermore, we introduce a new algorithm for the automated determination of inversion parameters. 
To put it concisely, this algorithm works by selecting the inversion parameters that maximize the agreement in the inversion result from different reference models. 
We apply this technique to the areas where the limited set of observed asteroseismic modes have resolving power, i.e.~in the interior 30\% of the star. 
We first demonstrate the efficacy of the algorithm by inverting the frequency differences between known models to determine that we are capable of producing the correct result. 
We then apply the method to the solar-type components of the 16~Cyg system with data obtained from the \emph{Kepler} mission.

\vfill\null
\Needspace{3\baselineskip}
\section{Methods}
We seek to measure the difference in internal structure between stars and their best-fitting evolutionary models, which we assume to be sufficiently close in structure such that linear perturbation theory applies. 
We begin by explicitly expanding Equation (\ref{eq:inversion}) using the $(u', Y)$ kernel pair. 
Given a set of $\mathscr{M}$ pulsation modes whose frequencies $\boldsymbol\nu$ have been measured, e.g.~$\mathscr{M}=\left\{(\ell=0, n=10), (\ell=1, n=12), \ldots\right\}$, for each mode of oscillation $i\in\mathscr{M}$ we have an equation relating a frequency perturbation to perturbations in stellar structure: 
\ifhbonecolumn
\begin{equation} \label{eq:forward} 
         \frac{\delta\nu_i'}{\nu_i'} 
         = 
          \int \KuY(r) \cdot \dun(r) \; \text{d}r 
         +\int \KYu(r) \cdot \dY(r)  \; \text{d}r 
         +\frac{F_{\text{surf}}(\nu_i')}{\nu_i'\cdot \mathcal{I}_i} 
         +\epsilon_i. 
\end{equation} 
\else
\begin{align} \label{eq:forward} 
         \frac{\delta\nu_i'}{\nu_i'} 
         = 
         &\int \KuY(r) \cdot \dun(r) \; \text{d}r 
\notag\\+&\int \KYu(r) \cdot \dY(r)  \; \text{d}r 
         +\frac{F_{\text{surf}}(\nu_i')}{\nu_i'\cdot \mathcal{I}_i} 
         +\epsilon_i. 
\end{align} 
\fi
Here $\delta \nu'$ is the difference in dimensionless oscillation mode frequency in the sense of (model - star), 
$\delta u'(r)$ is the difference in the dimensionless squared isothermal sound speed between a given stellar model and the star at fractional radius $r$, 
and $\dY(r)$ is the difference in the helium abundance. 
We assume the unknown differences between the true and the measured frequencies $\boldsymbol\epsilon$ to be independent and normally distributed with zero mean and known standard deviations $\boldsymbol{\sigma}$. 
The kernel functions $\boldsymbol{K}^{(u', Y)}$ and $\boldsymbol{K}^{(Y, u')}$ are known functions of the reference model and serve to relate changes in $u'$ and $Y$ to changes in oscillation mode frequencies. 
Finally, $F_{\text{surf}}$ is a surface term that depends on frequency and is normalized by mode inertiae $\boldsymbol{\mathcal{I}}$. Here we use the BG14-2 surface term, which \citet{2015ApJ...808..123S} showed to be a good choice. This relation has 
\begin{equation}
    F_{\text{surf}}(\nu'; \nu'_{\text{ac}}, \mathbf a) = 
           a_1 \left( \frac{\nu'}{\nu_{\text{ac}}'} \right)^{-1} 
         + a_2 \left( \frac{\nu'}{\nu_{\text{ac}}'} \right)^3 
\end{equation}
where $\mathbf a$ are coefficients that must be estimated during the inversion procedure and $\nu_{\text{ac}}'$ is the dimensionless acoustic frequency cut-off, which, under assumption of ideal gas, can be approximated \mb{by scaling from solar values with \citep{1991ApJ...368..599B}} 
\begin{equation}
    \nu_{\text{ac}}' 
    = 
    \nu_{\text{ac},\odot} \cdot
    \frac{g}{g_{\odot}} 
    \left( 
        \frac{T_{\text{eff}}}{T_{\text{eff},\odot}} 
    \right)^{-1/2} 
    \left( 
        \frac{R^3}{G M} 
    \right)^{1/2}
\end{equation}
with $g$ being the surface gravity of the reference model, $T_{\text{eff}}$ its effective temperature, $G$ the gravitational constant, \mb{and quantities subscripted with $\odot$ indicating the solar value.} 
The next step is to invert Equation (\ref{eq:forward}) to infer $\delta u'/u'(r)$, for which we will use the OLA technique. 

\Needspace{5\baselineskip}
\subsection{Optimally Localized Averages (OLA)}
We invert Equation (\ref{eq:forward}) using the method of Optimally Localized Averages (OLA).
If, for the sake of argument, the $(u',Y)$ kernel function of an oscillation mode were a $\delta$ function located at $r_0$ and zero elsewhere, and also if the $(Y,u')$ kernel were zero everywhere, then a departure in frequency of this mode from the observed value would demand that $u'(r_0)$ differs between model and star. 
According to Equation (\ref{eq:forward}), the relative difference in $u'(r_0)$ between the model and the star would be proportional to the relative difference in that mode's frequency. The OLA inversion technique works based on this concept. 

OLA combines the kernels of the observed modes into an \emph{averaging~kernel}~$\mathscr{K}$ resembling a localized function which is peaked at a chosen target radius inside the star. 
This is done via a linear combination of Equation (\ref{eq:forward}) over the observed modes, where each mode $i \in \mathscr{M}$ is weighted by a coefficient $c_i$. 
If a vector of coefficients $\mathbf c$ exists such that an averaging kernel with the desired properties can be formed, the inversion result, i.e., the relative difference in $u'$ between the model and the star, is then given by that same combination of the data. 
The process that creates the averaging kernel for $u'$ also combines the kernels of $Y$ to create a \emph{cross-term~kernel},~$\mathscr{C}$, and a reliable inversion result depends on $\mathscr{C}$ being as small as possible. 
Under these conditions, and assuming the surface term has been removed, the inversion result corresponds to an average of the underlying true difference weighted by the averaging kernel, i.e.,
\begin{equation}
    \left\langle\dun\right\rangle(r_0)
    =
    \int{\mathscr{K}}(r,r_0)\cdot \dun(r) \; \text{d}r
\end{equation}
assuming that $\int{\mathscr{K}}\;\text{d}r=1$.
Of course, the influence of data uncertainties must  be controlled as well.

More formally, for a given target radius $r_0$, the OLA procedure aims to construct an averaging kernel $\mathscr{K}(r)$ that is well-localized around $r=r_0$. 
Recalling Equation (\ref{eq:forward}), OLA proceeds by constructing a linear combination over all the observed modes: 
\begin{align} \label{eq:OLA}
    \sum_{i \in \mathscr{M}} c_i(r_0) \frac{\delta\nu_i'}{\nu_i'}
    =
    &\int \mathscr{K}(r; r_0, \mathbf c) \cdot \dun(r) \; \text{d}r \notag
\\ +&\int \mathscr{C}(r; r_0, \mathbf c) \cdot \dY(r) \; \text{d}r \notag
\\ +&\sum_{i \in \mathscr{M}} c_i(r_0) \cdot F_{\text{surf}}(\nu_i'; \nu'_{\text{ac}}, \mathbf a)/\left(\nu_i' \cdot \mathcal{I}_i \right) \notag
\\ +&\sum_{i \in \mathscr{M}} c_i(r_0) \cdot \epsilon_i
\end{align}
where the vector $\mathbf c$ are inversion coefficients that will need to be determined for each given $r_0$ and
\begin{align}
    \mathscr{K}(r; r_0, \mathbf c) &= \sum_{i \in \mathscr{M}} c_i(r_0) \cdot K_i^{(u, Y)}(r)
\\  \mathscr{C}(r; r_0, \mathbf c) &= \sum_{i \in \mathscr{M}} c_i(r_0) \cdot K_i^{(Y, u)}(r)
\end{align}
subject to the constraint that 
\begin{equation} \label{eq:sum-to-one}
    \int \mathscr{K}(r; r_0) \; \text{d}r = 1.
\end{equation}
Provided that the averaging kernel is well-localized at the target radius and the cross-term kernel, the surface-term contributions, and the combined data uncertainties are all small; this combination of relative frequency differences gives a localized average of $\delta u'/u'$ at the target radius $r_0$:
\begin{equation} \label{eq:local-avg}
    \left\langle \dun \right\rangle (r_0) = \sum_{i \in \mathscr{M}} \left( c_i(r_0) \cdot \frac{\delta\nu_i'}{\nu_i'} \right). 
\end{equation}
Here we have chosen to express relative differences in the sense
\begin{equation} \label{eq:rel-diff}
    \frac{\delta q}{q} = \frac{(\text{model} - \text{star})}{\text{model}} = \frac{ (q_{\text{ref}} - q_{\text{star}}) }{q_{\text{ref}}}
\end{equation} 
where $q$ can refer to any quantity. 
Thus, Equation~(\ref{eq:local-avg}) can be re-dimensionalized using Equation~(\ref{eq:MR}) to infer $u_{\text{star}}$ with
\begin{equation} \label{eq:dimensional}
     u_{\text{star}}(r) = \left( 1 - \dun(r) + \dRR - \dMM \right) \cdot u_{\text{ref}}(r).
\end{equation}
We now turn our attention to determining the coefficients $\mathbf c$ that make this estimate possible.

\Needspace{3\baselineskip}
\subsection{Inversion coefficients using Subtractive OLA}
The optimal inversion coefficients $\boldsymbol {\hat c}$ must strike a balance between forming a well-localized averaging kernel and forming a small cross-term kernel while still having small uncertainty. 
In Subtractive OLA \citep[SOLA,][]{1992A&A...262L..33P, 1994A&A...281..231P}, the averaging kernel is formed according to a specified well-localized form (the ``target~kernel''), and the coefficients $\mathbf c$ are determined by minimizing the difference between the averaging kernel obtained and the target kernel. 
This is a fast implementation of the OLA method. 
It comes at the price of a free parameter in the form of the properties of the target kernel. 
SOLA determines optimal coefficients $\mathbf{\hat{c}}$ for a given target radius $r_0$ by solving the optimization problem 
\ifhbonecolumn
\begin{align} \label{eq:opt}
        \boldsymbol {\hat c}(r_0; \beta, \mu, \Delta)
        =
        {}&\underset{\mathbf{c}}{\arg\min} \; \Bigg\{
            \mathscr{F}(\mathbf c; r_0, \Delta) 
          + \beta \int \mathscr{C}(r; r_0, \mathbf c)^2 \; \text{d}r 
          + \mu \sum_{i\in\mathscr{M}} \left( c_i^2 \cdot \sigma_i^2 \right)
         \Bigg\} 
         \notag\\\text{subject~to} & \; 
        \int{\mathscr{K}(r; r_0, \mathbf{c})} \; \text{d}r = 1
\quad\text{ and }\quad  \sum_{i\in\mathscr{M}} c_i \cdot \frac{F_{\text{surf}}(\nu'_i; \nu_{\text{ac}})}{\nu'_i \cdot \mathcal{I}_i} = 0.
\end{align}
\else
\begin{align} \label{eq:opt}
\boldsymbol {\hat c}(r_0; \beta, \mu, \Delta)
        =
        {}\underset{\mathbf{c}}{\arg\min} \; \Bigg\{
            &\mathscr{F}(\mathbf c; r_0, \Delta) 
    \notag\\+{}\beta \int \mathscr{C}(r; r_0, \mathbf c)^2 \; \text{d}r 
            +{}&{} \mu \sum_{i\in\mathscr{M}} \left( c_i^2 \cdot \sigma_i^2 \right)
         \Bigg\} 
         \notag\\
     \text{subject~to} \; 
        \int \mathscr{K}(r; r_0, \mathbf{c}) \;\text{d}r &= 1
\notag\\\text{and}\; \sum_{i\in\mathscr{M}} c_i \cdot \frac{F_{\text{surf}}(\nu'_{i}; \nu_{\text{ac}})}{\nu'_i \cdot \mathcal{I}_{i}} &= 0.
\end{align}
\fi
Here $\beta$ and $\mu$ are parameters that must be chosen to penalize the amplitude of the cross-term kernel and the effect of data uncertainties, respectively. 
A third parameter, $\Delta$, gives the width of the target kernel(s). 
The function $\mathscr{F}$ penalizes deviations of the averaging kernel from the target~kernel~$\mathcal{T}$ and can be calculated as
\begin{equation}
        \mathscr{F}(\mathbf c; r_0, \Delta)
        =  
        \int \left[ \mathscr{K}(r; r_0, \mathbf c) - \mathcal{T}(r; r_0, \Delta) \right]^2 \; \text{d}r. 
\end{equation}
The functional form of $\mathcal{T}$ can be chosen e.g.~as a modified Gaussian that decays to zero at $r=0$ but remains peaked at $r=r_0$ \citep[e.g.][]{1999MNRAS.309...35R} with 
\begin{align}
    \mathcal{T}(r; r_0, \Delta) &= A\cdot r\cdot \exp\left\{-\mathcal{G}(r; r_0, \Delta)^2\right\} \\
    \mathcal{G}(r; r_0, \Delta) &= \frac{r-r_0}{D(r_0, \Delta)} + \frac{D(r_0, \Delta)}{2 r_0}.
\end{align}
The normalization factor $A$ is chosen to ensure $\int \mathcal{T} \; \text{d}r = 1$. 
Since the resolution ultimately depends on the internal sound speed $c_s$ \citep[][]{1993ASPC...42..141T}, the function $D$ gives the width of the kernels according to variations in $c_s$ and a free parameter $\Delta$ that describes a fiducial width as 
\begin{align}
    D(r_0, \Delta) &= \Delta \cdot \frac{c_s(r_0)}{c_s(r_f)} 
\end{align}
with $r_f$ being an arbitrary reference point (e.g. we choose $r_f=0.2$, although the result is rather insensitive to the choice). 
We note that other choices of $\mathscr{F}$, $\mathcal{T}$, $\mathcal{G}$, and $D$ are possible \citep[see e.g.][]{1985SoPh..100...65G, 1989ApJ...343..526B}, but they will not be explored here. 

The SOLA inversion problem can be cast into a system of linear equations with the constraints enforced using Lagrange multipliers. 
Given choices of $\beta, \mu,$ and $\Delta$, Equation (\ref{eq:opt}) can be solved via matrix inversion, the details of which can be found for example in Chapter 10 of \citealt{BasuChaplin2017}. 
See \citet{1999MNRAS.309...35R} for a description of how inversion parameters are usually selected in helioseismology. 
Depending on the data that are available, it may be possible to form zero, one, or more well-localized averaging kernels with correspondingly small cross-term kernels and well-controlled uncertainties at different locations in the stellar interior.

\Needspace{3\baselineskip}
\subsection{Selecting inversion parameters with multiple reference models (``inversions-for-agreement'')} 
\label{sec:inversion-for-agreement}
It is not clear \emph{a priori} which inversion parameters should be chosen, nor is there a reliable algorithm for their selection. 
Here we propose an algorithm for selecting inversion parameters based on the following information. 
First, beside the effects that stem from differences in $M$ and $R$, inversion results do not otherwise depend on the choice of reference model: with proper selection of inversion parameters, a wide range of reference models are capable of producing the correct inference \citep{2000ApJ...529.1084B}. 
\mb{Furthermore, for a given mode set, and setting aside the surface term, the values of the mode frequencies themselves do not play a role in determining the averaging and cross-term kernels.} 
Thus, provided the differences in the kernels between models are small, the same inversion parameters can be used for different models. 
Instead of performing single-model inversions, we invert using an array of reference models that span the uncertainties in $M$ and $R$. 
We simultaneously estimate the inversion parameters and the stellar $M$ and $R$ such that the inferred stellar $u$ profile from the different models are in agreement. 
We achieve this via repeated iterative optimization with random noise realizations. 
We constrain $M$ and $R$ with normal priors based on past studies, and set uniform priors on the inversion parameters. 
\mb{We have also tried this procedure with each reference model having its own individual set of inversion parameters ($\beta, \mu, \Delta$) to optimize, and we found that it did not have a substantial impact on the results. }

We generate an array of \mb{9} reference models that are calibrated to span the $1\sigma$ uncertainties in mass and radius for each star whose interior structure we seek to infer. 
We optimize a vector of 5 inversion parameters 
$\boldsymbol\alpha = (\beta, \mu, \Delta, M_{\text{star}}, R_{\text{star}})$ 
which are shared among the \mb{9} models. 
We take an average among their inferred values of $u_{\text{star}}$, and finally we choose the $\boldsymbol \alpha$ that minimizes the variance of this average, weighted by the priors on $M_{\text{star}}$ and $R_{\text{star}}$. 
Formally, we postulate that the optimal inversion parameters $\boldsymbol{\hat{\alpha}}$ across all of the reference models is 
\begin{align} \label{eq:invert-for-agree}
    \boldsymbol{\hat{\alpha}}
    =
    \underset{\boldsymbol{\alpha}}{\arg\min} \Bigg\{ 
    &\sum_{r_j\in \boldsymbol{r_0}} 
            \log \text{Var} \left[ 
                \tilde u\left(r_j; \boldsymbol {\alpha} \right) 
            \right] - \log \Psi(\boldsymbol\alpha)
    \Bigg\}
\end{align}
where 
$\text{Var}$ is the variance operator, 
$\mathbf{r_0}$ are the target radii, 
and $\tilde u$ is a vector whose $k^{\text{th}}$ element 
$u_k(r_0; \boldsymbol \alpha)$ gives the inferred value of $u_{\text{star}}$ at target radius $r_0$ via the $k^{\text{th}}$ reference model using the inversion parameters $\boldsymbol \alpha$
(\emph{cf}.~Equations.~\ref{eq:local-avg}-\ref{eq:opt}). 
Finally, $\Psi$ is the prior distribution, which in this case has
\begin{equation}
    \Psi(\boldsymbol \alpha)
    =
    \psi\left(M_{\text{star}}; \mu_M, \sigma^2_M\right) \cdot
    \psi\left(R_{\text{star}}; \mu_R, \sigma^2_R\right)
\end{equation}
with $\psi$ being the normal density function and 
$\mu_x$ and $\sigma_x$ being the mean and standard deviation of $x$. 
In each iteration of the algorithm, each of the non- and re-dimensionalizations are performed with the current estimate of $M_{\text{star}}$ and $R_{\text{star}}$. 
For example,
\ifhbonecolumn
\begin{equation}
    \frac{\delta \nu'}{\nu'}
    =
    \left[
        \left(
            \frac{R_{\text{ref}}^{3/2}}{M_{\text{ref}}^{1/2}}
        \right) 
        \nu_{\text{ref}}
        -
        \left(
            \frac{R_{\text{star}}^{3/2}}{M_{\text{star}}^{1/2}}
        \right)
        \nu_{\text{star}}
    \right]
    / 
    \left[
        \left(
            \frac{R_{\text{ref}}^{3/2}}{M_{\text{ref}}^{1/2}}
        \right)
        \nu_{\text{ref}}
    \right]. 
\end{equation} 
\else
\begin{gather}
    \frac{\delta \nu'}{\nu'}
    =
    \left[
        \left(
            \frac{R_{\text{ref}}^{3/2}}{M_{\text{ref}}^{1/2}}
        \right) 
        \nu_{\text{ref}}
        -
        \left(
            \frac{R_{\text{star}}^{3/2}}{M_{\text{star}}^{1/2}}
        \right)
        \nu_{\text{star}}
    \right]
    / \notag \\
    \left[
        \left(
            \frac{R_{\text{ref}}^{3/2}}{M_{\text{ref}}^{1/2}}
        \right)
        \nu_{\text{ref}}
    \right]. 
\end{gather} 
\fi
In summary, Equation (\ref{eq:invert-for-agree}) says that the optimal inversion parameters are the ones that give the same inference of $u_{\text{star}}$ across all the reference models.

Since the inversion results depend on uncertain measurements, we perform repeated trials with random realizations of noise. 
Specifically, in each trial, we perturb each frequency $\nu$ with normal noise according its uncertainty $\sigma_{\nu}$, and the mass and radius estimates $\mu_M$ and $\mu_R$ via their uncertainties $\sigma_M$ and $\sigma_R$. 
We then use the \citet{nelder1965simplex} downhill simplex method to numerically search for the parameters that satisfy Equation (\ref{eq:invert-for-agree}) for that realization of noise. 
Because each inversion parameter is strictly non-negative and can potentially take on a large range of values, we optimize $\log \boldsymbol \alpha$. 
We stop each trial after either the relative change in the objective function is reduced by less than the square root of the machine precision for double precision floating point numbers ($\sim 10^{-8}$), or a maximum number of 512 iterations is reached. 
In the majority of cases, the former condition is met. 
We perform 128 trials and report the averaged results. 
Finally, we visually inspect the resulting averaging kernels and cross-term kernels to ensure that the averaging kernels are well-localized at the target radii and that the cross-term kernels have small amplitude everywhere.

\Needspace{5\baselineskip}
\section{Results} 
\subsection{Tests on Models} 
In order to validate our technique, we first apply the method to known models; this allows us to check that the procedure does indeed produce the correct result.
Specifically, we determine whether or not we can accurately recover the internal $u$ profiles of the \emph{GOE} models of 16~Cyg~A and B using an array of different reference models as reference.

For the test, we generate an array of reference models for each star by calibrating models to their estimated masses ($\pm 1 \sigma$, \citealt{2016ApJ...830...31B}), radii ($\pm 1 \sigma$, \citealt{2013MNRAS.433.1262W}), ages \citep{2016ApJ...830...31B}, luminosities \citep{2013MNRAS.433.1262W}, and metallicities \citep{2009A&A...508L..17R}. 
The estimates we use for these stars are given in Table~\ref{tab:stellar-parameters}. 
\mb{We calculate the models using the given mean values of their ages, luminosities, and metallicities.} 
We construct the models using the MESA stellar evolution code \citep[\emph{Modules for Experiments in Stellar Astrophysics},][]{2011ApJS..192....3P}. 
For each model, we use ADIPLS \citep[\emph{the Aarhus adiabatic oscillation package},][]{2008ApSS.316..113C} to calculate the adiabatic oscillation mode frequencies corresponding to the 54 and 56 oscillation modes that have been identified in 16~Cyg~A and B, respectively. 
We use the same treatments of evolution and pulsation that are described in Section 2.1 of \citealt{2016ApJ...830...31B}. 
None of the reference models have exactly the same mass or radius as the two \emph{GOE} models that we are treating as our proxy stars. 
We perturb the proxy star frequencies with noise prior to beginning the procedure. 

\ifhbonecolumn\singlespace\fi
\startlongtable
\begin{deluxetable*}{llcccccccc}
\tablecaption{Fundamental parameters of 16~Cyg~A and B. 
\label{tab:stellar-parameters}}
\tablecolumns{8}
\tablenum{1}
\tablewidth{0pt}
\tablehead{
    \colhead{Name} &
    \colhead{Mass} &
    \colhead{Radius} &
    \colhead{Age} &
    \colhead{Luminosity} &
    \colhead{Metallicity} &
    \\
    \colhead{} &
    \colhead{$\mu_M \pm \sigma_M$} &
    \colhead{$\mu_R \pm \sigma_R$} &
    \colhead{$\tau$} &
    \colhead{$L$} &
    \colhead{$[$Fe$/$H$]$} &
    \\
    \colhead{} &
    \colhead{$[\text{M}_\odot]$} &
    \colhead{$[\text{R}_\odot]$} &
    \colhead{$[\text{Gyr}]$} &
    \colhead{$[\text{L}_\odot]$} &
    \colhead{(dex)} &
}
\startdata 
    16~Cyg~A & 1.080 $\pm$ 0.016 & 1.22 $\pm$ 0.02 & 6.90 $\pm$ 0.40 & 1.56 $\pm$ 0.05 & 0.096 $\pm$ 0.026 \\ 
    16~Cyg~B & 1.030 $\pm$ 0.015 & 1.12 $\pm$ 0.02 & 6.80 $\pm$ 0.28 & 1.27 $\pm$ 0.04 & 0.052 $\pm$ 0.021 \\ 
\enddata 
\end{deluxetable*}
\ifhbonecolumn\doublespace\fi

\begin{figure*}
    \centering
    \captionsetup{width=\captwidth,font=small}
    \makebox[\colwidth][c]{%
        \adjustbox{trim={0 \bottrim cm \righttrim cm 0.1cm},clip}{
            \includegraphics[width=\thinfigstar]{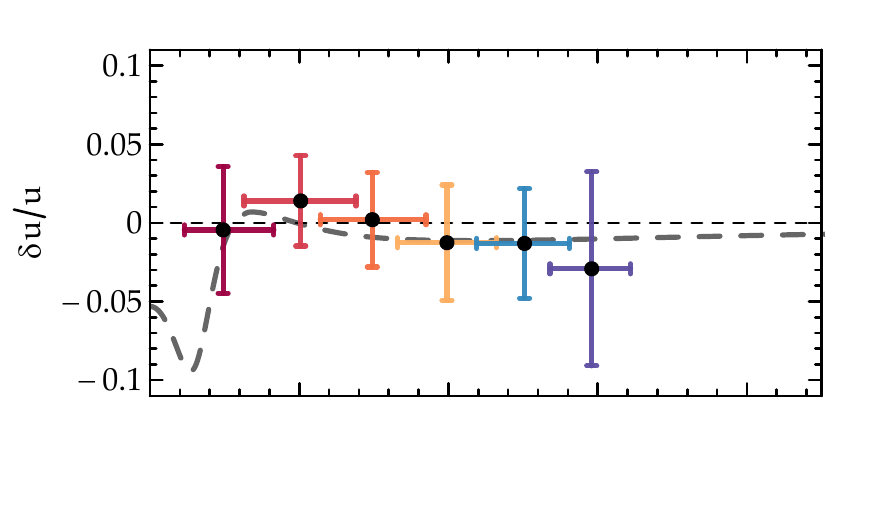}
        }%
        \adjustbox{trim={1.45cm \bottrim cm 0 0.1cm},clip}{
            \includegraphics[width=\thinfigstar]{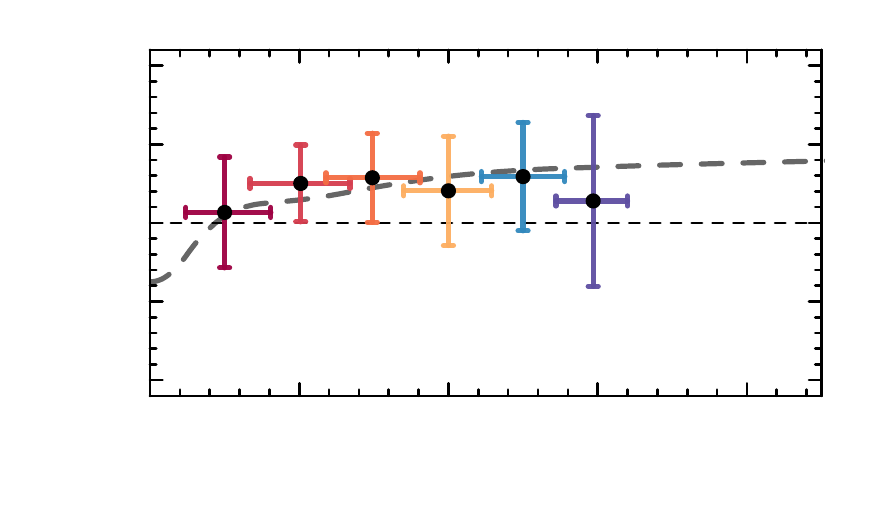}
        }
    }\\
    \makebox[\colwidth][c]{%
        \adjustbox{trim={0 \bottrim cm \righttrim cm 0.25cm},clip}{
            \includegraphics[width=\thinfigstar]{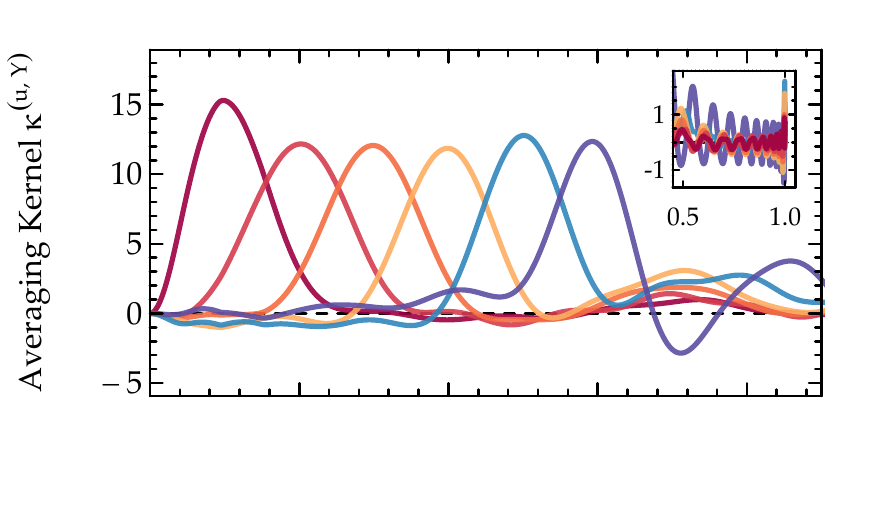}%
        }
        \adjustbox{trim={1.45cm \bottrim cm 0 0.25cm},clip}{
            \includegraphics[width=\thinfigstar]{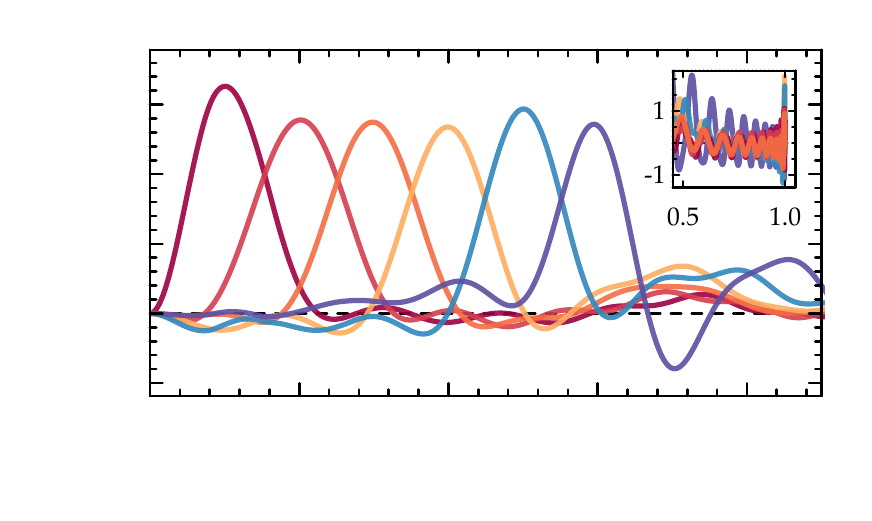}
        }
    }\\
    \makebox[\colwidth][c]{%
        \adjustbox{trim={0 0 \righttrim cm 0.25cm},clip}{
            \includegraphics[width=\thinfigstar]{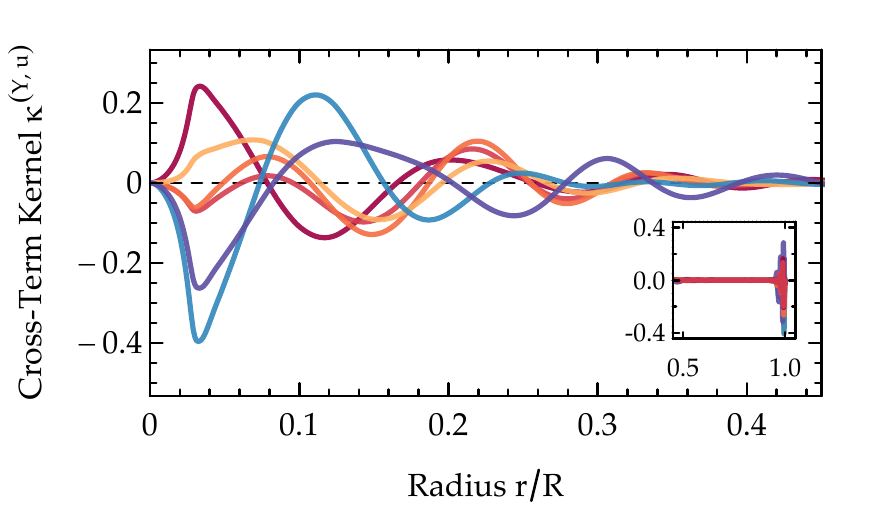}
        }%
        \adjustbox{trim={1.45cm 0 0 0.25cm},clip}{
            \includegraphics[width=\thinfigstar]{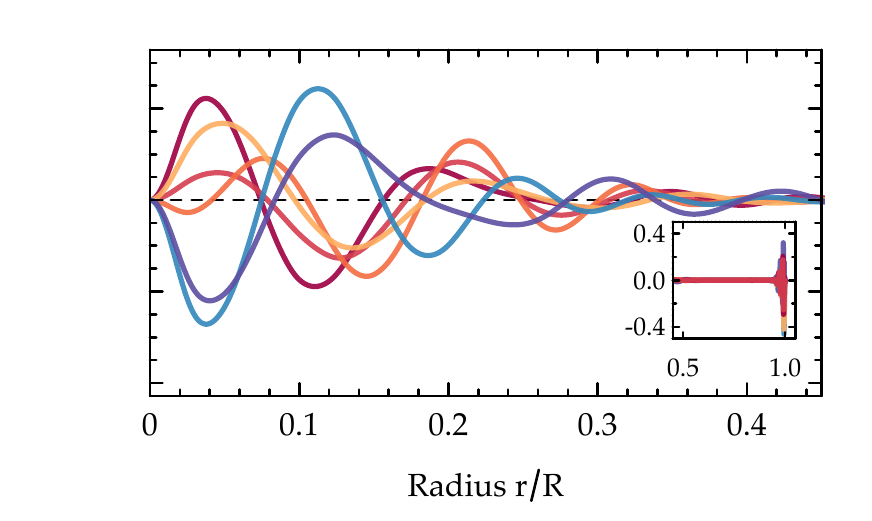}
        }
    }
    \caption{ Structural inversions for the internal squared isothermal sound-speed profile $u$ of evolutionary models of 16~Cyg~A (left) and 16~Cyg~B (right). 
    \textbf{Top}: actual relative difference $\delta u/u$ between the evolutionary model and \mb{a reference model from the corresponding array of reference models for that star} (dashed gray line), and the result of the inversion-for-agreement procedure presented here (colored points). 
    The colors serve to associate the inversion results with their respective averaging and cross-term kernels. 
    \textbf{Middle}: averaged averaging kernels, sensitive to changes in $u'$, which have been placed at target radii $\mathbf{r_0} = [0.05, 0.1, 0.15, 0.2, 0.25, 0.3]$. 
    \textbf{Bottom}: averaged cross-term kernels that are sensitive to changes in helium abundance, whose amplitudes should be small everywhere relative to the averaging kernels. 
    \textbf{Insets}: the behavior of the averaging and cross-term kernels closer to the surface, where their amplitudes are small as desired (note the change in axes). 
    \label{fig:model-test} }
\end{figure*}

We apply the inversion-for-agreement procedure described in Section~\ref{sec:inversion-for-agreement}. 
The results are shown in Figure~\ref{fig:model-test}. 
The procedure gets the correct result. 
The uncertainties in $\delta u/u$ are given by the average over the 128 trials. 
The ``uncertainties'' in fractional radius $r/R$ are a measure of the resolution of the inversion and are given by the width at half maximum of an average over the averaging kernels of the different trials. 
The averaging kernels are reasonably well-localized and the cross-term kernels are small everywhere. 
The averaging kernels placed at $r_0=0.3$ begin to develop some amplitude outside of target region; this is why we do not attempt to probe shallower layers.

\Needspace{3\baselineskip}
\subsection{Inversions for stellar structure} 
\begin{figure*}
    \centering
    \captionsetup{width=\captwidth,font=small}
    \makebox[\colwidth][c]{%
        \adjustbox{trim={0 0 \righttrim cm 0.1cm},clip}{
            \includegraphics[width=\thinfigstar]{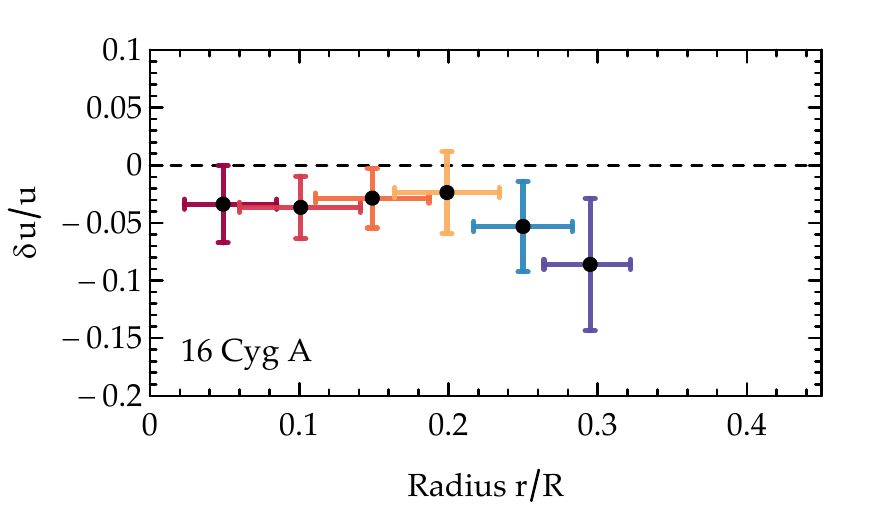}
        }%
        \adjustbox{trim={1.45cm 0 0 0.1cm},clip}{
            \includegraphics[width=\thinfigstar]{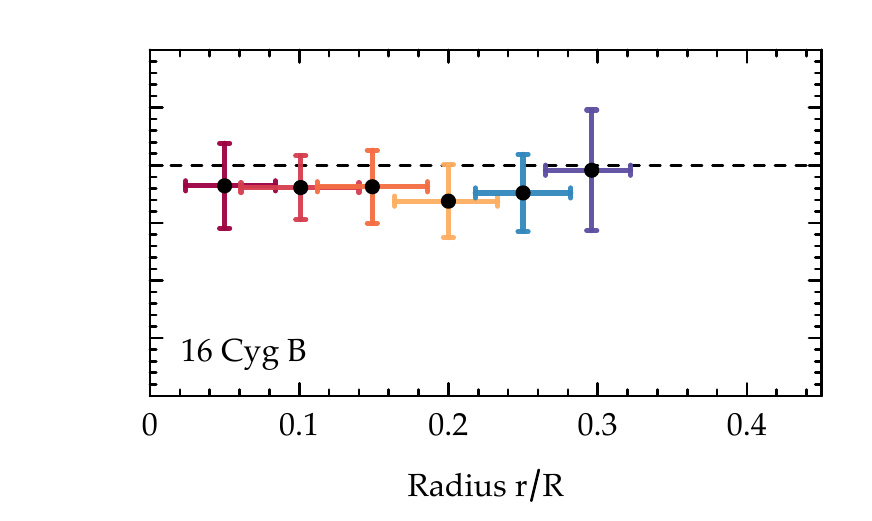}
        }
    }
    \caption{ Structural inversions for the internal squared isothermal sound-speed profile $u$ of 16~Cyg~A (left) and 16~Cyg~B (right) using the inversion-for-agreement technique introduced in this paper. 
    Results are shown in terms of relative differences with respect to the \emph{GOE} evolutionary models of these stars (\emph{cf}.~Equation~\ref{eq:rel-diff}). 
    The sound speeds in the cores of 16~Cyg~A \& B are greater than those of the evolutionary models. 
    \label{fig:Cyg-inversions} 
    }
\end{figure*}

We now apply our structure inversion-for-agreement procedure on asteroseismic data of 16~Cyg~A and B. 
The relative differences with respect to the \emph{GOE} evolutionary models of these stars are shown in Figure~\ref{fig:Cyg-inversions}. 
As the mode sets are the same as in our tests with models, the averaging kernels and cross-term kernels are nearly identical to those shown in Figure~\ref{fig:model-test}.  
The results are also tabulated in Tables~\ref{tab:CygA}~and~\ref{tab:CygB}. 
We find that the sound speeds throughout the cores of 16~Cyg~A and B exceed those of these evolutionary models.

\mb{In the case of 16~Cyg~A, each of the individual measurements hovers around a $1\sigma$ difference. 
On the one hand, all of the model sound speeds are found to be lower than in the star, indicating that there are systematic differences between the model and the star. 
Viewed this way, the overall result is more significant than each of the measurements taken separately. 
On the other hand, there is covariance between the different measurements, because the different averaging kernels overlap to some degree. 
Thus, assigning an overall level of statistical significance to these results is challenging. 
}

\ifhbonecolumn\else\clearpage\fi

\begin{figure*}
    \centering
    \captionsetup{width=\captwidth,font=small}
    \makebox[\colwidth][c]{%
        \adjustbox{trim={0 0 \righttrim cm 0.1cm},clip}{
            \includegraphics[width=\thinfigstar]{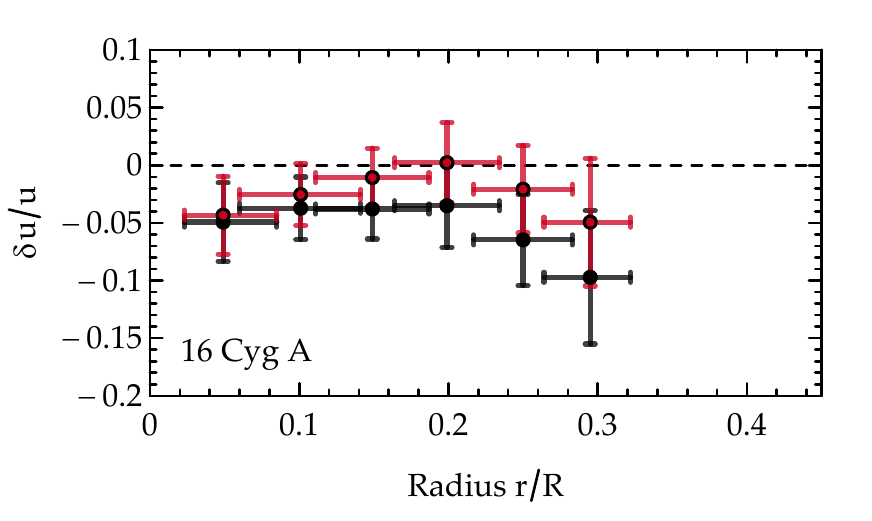}
        }%
        \adjustbox{trim={1.45cm 0 0 0.1cm},clip}{
            \includegraphics[width=\thinfigstar]{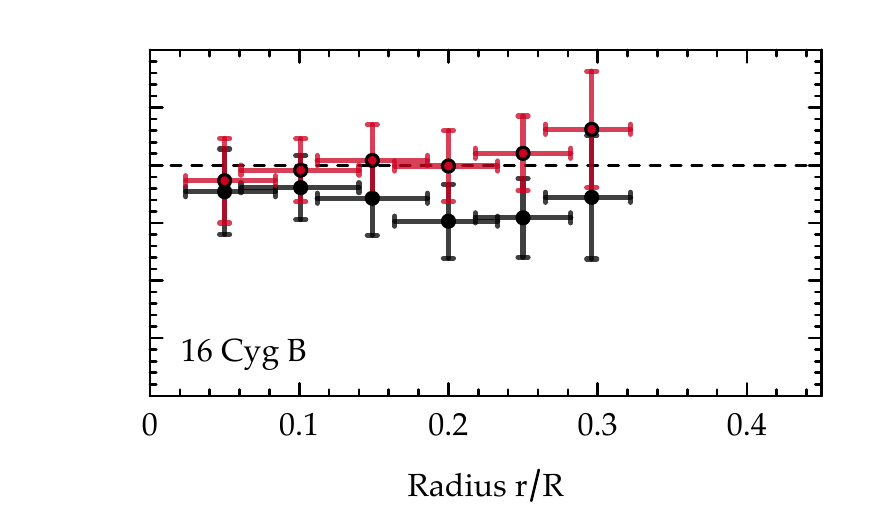}
        }
    }
    \caption{Relative differences between the inferred sound speeds $u$ of 16~Cyg~A (left) and 16~Cyg~B (right) shown against a model with a high radius and low mass (black points) and a model with a low radius and high mass (\textcolor{diff-red}{red} points). 
    \label{fig:ref-mods} }
\end{figure*}

To assess whether the differences may stem from the \emph{GOE} models having wrong masses or radii, we compare the inversion results against other models of different mass and radius. 
Following Equation (\ref{eq:MR}), the spread in sound speeds caused by mass and radius estimates are largest for the models with either a high radius and a low mass, or models with a low radius and high mass. 
Thus, we show in Figure~\ref{fig:ref-mods} these inversion results against models with masses and radii that differ by $1\sigma$ in opposite directions from the mean estimated masses and radii of these stars. 
In both cases, the models with higher masses and lower radii are preferred. 
However, while the 16~Cyg~B models show roughly broad agreement, the 16~Cyg~A models do not agree quite as well. 

\begin{figure}
    \centering
    \captionsetup{width=\captwidth,font=small}
    \makebox[\linewidth][c]{%
        \includegraphics[width=\figfactor\linewidth]{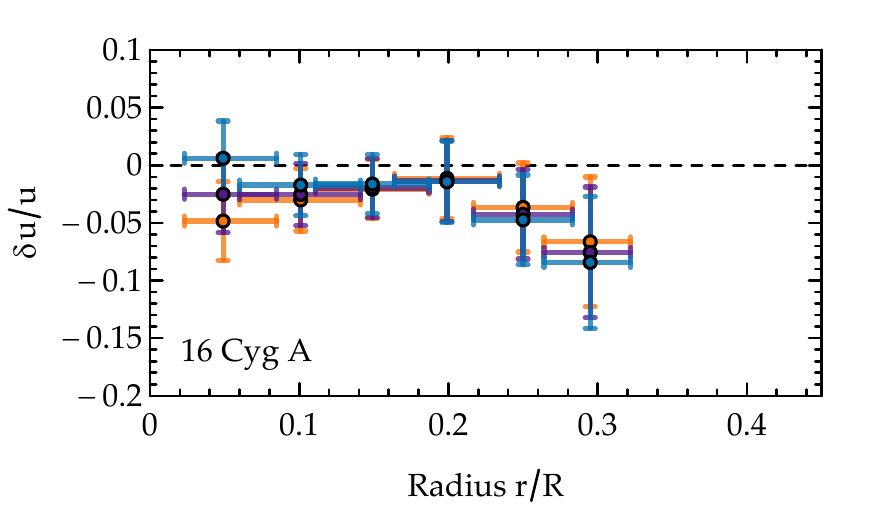}
    }
    \caption{Relative differences between the isothermal sound speed $u$ in the core of 16~Cyg~A and models with lower ages (6~Gyr in \textcolor{diff-purple}{purple}, 5~Gyr in \textcolor{diff-blue}{blue}). 
    A model of 16~Cyg~A at the mean estimated present age (6.9~Gyr in \textcolor{diff-orange}{orange}) is shown for reference. 
    \label{fig:low-age} }
\end{figure}

\begin{figure}
    \centering
    \captionsetup{width=\captwidth,font=small}
    \makebox[\linewidth][c]{%
        \includegraphics[width=\figfactor\linewidth]{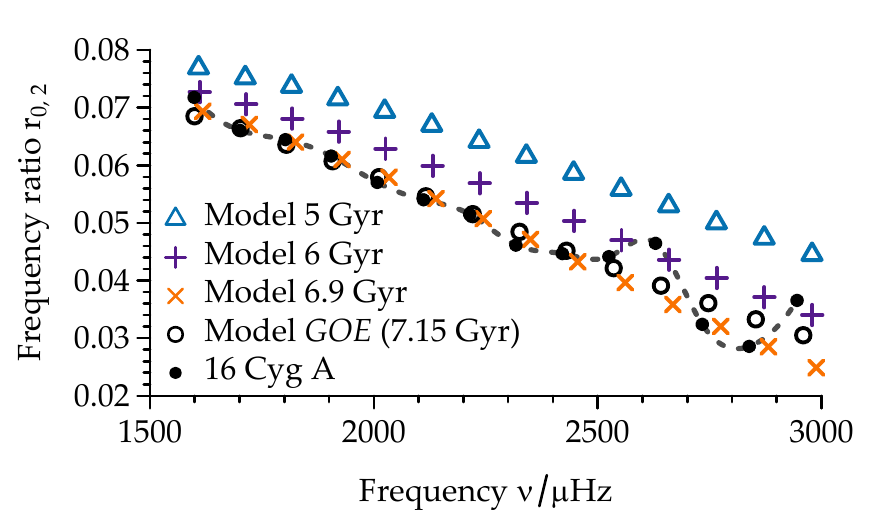}
    }
    \caption{The ratio of the small frequency separation to the large frequency separation---a core-conditions indicator that is insensitive to surface effects---against mode frequency for asteroseismic data of 16~Cyg~A in comparison with models of various ages. 
    \label{fig:r02} }
\end{figure}

The isothermal speed of sound depends principally on the inverse of the mean molecular weight $\mu$ of the fluid. 
Fusion alters the core composition and increases $\mu$; thus, with all else equal, older stars will have a lower $u$ in the core. 
To assess the effect of stellar age in the context of these results, we evolve two models to match the characteristics of 16~Cyg~A (\emph{cf}.~Table~\ref{tab:stellar-parameters}) with ages of $\tau=6$~Gyr and $\tau=5$~Gyr, which are significantly lower than the estimated age of $\tau=6.90\pm 0.40$~Gyr. 
The relative differences between the core $u$ of 16~Cyg~A and these models are shown in Figure~\ref{fig:low-age}. 
In the deep core ($r=0.05$), the young age models have smaller differences when compared with the \emph{GOE} model. 
However, the differences farther out are not explained with smaller ages. 
Furthermore, although it seems the inner core is better with the low-age models, frequency combinations such as $r_{0,2}$ \citep{2003A&A...411..215R} indicate that the low age models are not appropriate. 
A comparison of $r_{0,2}$ values for these models is shown in Figure~\ref{fig:r02}.
This may explain why the differences in $u$ worsen just outside the core.

\ifhbonecolumn\singlespace\fi
\startlongtable
\begin{deluxetable*}{ccccccc}
\tablecaption{Results of inversions for the squared isothermal sound speed $u$ inside of 16~Cyg~A at different target radii $r_0$ in the stellar core.
\label{tab:CygA}}
\tablecolumns{7}
\tablenum{2}
\tablewidth{0pt}
\ifhbonecolumn\tabletypesize{\small}\fi
\tablehead{
    \colhead{Target radius} & \colhead{Peak of $\mathscr{K}$}    & \colhead{Relative $u$ difference}          & \colhead{Sq.~iso.~sound~speed} \\
    \colhead{$r_{0}$}       & \colhead{$r_{\max} \pm \text{FWHM}$} & \colhead{$\delta u/u \pm \sigma_{\delta}$} & \colhead{$u \pm \sigma_u$} \\
    \colhead{[R]}           & \colhead{[R]}                      & \colhead{$[\text{w.r.t.~model~\emph{GOE}}]$}                 & \colhead{[$10^{15}$ cm$^2$ s$^{-2}$]}}
\startdata 
    0.05 & 0.049 $\pm$ 0.031 & -0.033 $\pm$ 0.033 & 1.515 $\pm$ 0.049 \\ 
    0.10 & 0.101 $\pm$ 0.041 & -0.036 $\pm$ 0.027 & 1.580 $\pm$ 0.041 \\ 
    0.15 & 0.149 $\pm$ 0.038 & -0.028 $\pm$ 0.025 & 1.404 $\pm$ 0.035 \\ 
    0.20 & 0.199 $\pm$ 0.035 & -0.023 $\pm$ 0.035 & 1.181 $\pm$ 0.041 \\ 
    0.25 & 0.250 $\pm$ 0.033 & -0.053 $\pm$ 0.039 & 1.019 $\pm$ 0.037 \\ 
    0.30 & 0.295 $\pm$ 0.029 & -0.086 $\pm$ 0.057 & 0.910 $\pm$ 0.048 \\ 
\enddata 
\end{deluxetable*}
\ifhbonecolumn\doublespace\fi

\ifhbonecolumn\singlespace\fi
\startlongtable
\begin{deluxetable*}{ccccccc}
\tablecaption{Results of inversions for the squared isothermal sound speed $u$ inside of 16~Cyg~B. 
\label{tab:CygB}}
\tablecolumns{7}
\tablenum{3}
\tablewidth{0pt}
\ifhbonecolumn\tabletypesize{\small}\fi
\tablehead{
    \colhead{Target radius} & \colhead{Peak of $\mathscr{K}$}    & \colhead{Relative $u$ difference}          & \colhead{Sq.~iso.~sound~speed} \\
    \colhead{$r_{0}$}       & \colhead{$r_{\max} \pm \text{FWHM}$} & \colhead{$\delta u/u$} & \colhead{$u \pm \sigma_u$} \\
    \colhead{[R]}           & \colhead{[R]}                      & \colhead{$[\text{w.r.t.~model~\emph{GOE}}]$}                 & \colhead{[$10^{15}$ cm$^2$ s$^{-2}$]}}
\startdata 
    0.05 & 0.050 $\pm$ 0.030 & -0.017 $\pm$ 0.036 & 1.485 $\pm$ 0.053 \\ 
    0.10 & 0.101 $\pm$ 0.039 & -0.019 $\pm$ 0.027 & 1.533 $\pm$ 0.041 \\ 
    0.15 & 0.149 $\pm$ 0.037 & -0.018 $\pm$ 0.031 & 1.402 $\pm$ 0.043 \\ 
    0.20 & 0.200 $\pm$ 0.034 & -0.031 $\pm$ 0.031 & 1.216 $\pm$ 0.037 \\ 
    0.25 & 0.250 $\pm$ 0.032 & -0.024 $\pm$ 0.033 & 1.025 $\pm$ 0.033 \\ 
    0.30 & 0.296 $\pm$ 0.028 & -0.004 $\pm$ 0.052 & 0.870 $\pm$ 0.045 \\
\enddata 
\end{deluxetable*}
\ifhbonecolumn\doublespace\fi

\Needspace{3\baselineskip}
\section{Discussion \& Conclusions} 
In this paper, we examined the problem of deducing the core structures of solar-like stars based on the frequencies of their normal modes of oscillation. 
We applied the SOLA inversion technique to infer the radial dependence of the squared isothermal sound speed throughout the interiors of two solar-type main-sequence stars. 
We inverted using the $(u',Y)$ kernel pair because the influence of the second variable ($Y$) is very low in the regions of our interest. 
We presented a new algorithm for the automated determination of inversion parameters that also accounts for imprecise/inaccurate stellar mass and radius estimates. 
We validated this technique on models, and then applied it to the well-studied stars 16~Cyg~A and B. 
We measured $u$ at several different radii within these stars and compared these values to best-fitting evolutionary models of these stars. 
We found that the sound speeds in the cores of these stars are greater than in the \emph{GOE} models. 
This is to our knowledge the first time the radial variation in sound speed has been measured in a star other than the Sun.

In the case of 16~Cyg~B, it seems plausible that adjustments to the mass and radius of the \emph{GOE} model may serve to fix the differences that we find. 
In the case of 16~Cyg~A however, the source of the disparities is more difficult to pinpoint. 
Lower age models help with the differences in the deeper parts of the core, but do not aid with the differences farther out. 
Furthermore, the lower age models fail to reproduce the asteroseismic frequency ratios of 16~Cyg~A, which effectively rules age out as the culprit. 
Missing physical processes, incorrect application of known processes, or inadequate inputs in the calculations of the models may therefore be at cause. 
\mb{For example, while the GOE model of 16~Cyg~A does not have a convective core at the present age, it did have one during the first 1.75~Gyr of its evolution. 
As core convection modifies the mean molecular weight, the duration of its existence may leave a footprint in the sound speed. 
It may then be the case that an incorrect prescription of convection in stellar cores is the cause of these discrepancies. }

16~Cyg~A and B are stars either on the main sequence or nearly at the main-sequence turnoff. 
The main sequence is a well-studied phase of evolution, and the different types of observations that are possible for main-sequence stars lead to estimates of their ages, masses, and radii in a well-known way. 
Being the first and also the longest-lived stage of evolution, getting the details of the main sequence evolution right is necessary for also getting the later stages of stellar evolution right as well. 
Any neglected processes that cause substantial errors on the core structure of main-sequence stars will subsequently propagate into the later stages of evolution.

As is always the case with ill-posed inverse problems, there is no guarantee that the end result will be the true profile of the star. 
That being said, the procedure has worked well in blind tests on models with known structure. 
Therefore, some confidence can be put in the results.

\Needspace{3\baselineskip}
\acknowledgments The research leading to the presented results has received funding from the European Research Council under the European Community's Seventh Framework Programme (FP7/2007-2013) / ERC grant agreement no 338251 (StellarAges). This research was undertaken in the context of the International Max Planck Research School for Solar System Research. E.P.B.\ acknowledges support from the National Physical Science Consortium Fellowship. S.B.\ acknowledges partial support from NSF grant AST-1514676 and NASA grant NNX13AE70G. \mb{We thank the anonymous referee for their very helpful report.} 

\software R 3.2.3 \citep{R}, magicaxis 2.0.0 \citep{magicaxis2}, 
kernel calculations by \citet{kerexact}, \mb{MESA \citep{2011ApJS..192....3P}, and ADIPLS \citep{2008ApSS.316..113C}.} 

\Needspace{3\baselineskip}
\bibliographystyle{aasjournal.bst}
\bibliography{astero}

\end{document}